\documentclass[a4paper]{article}

\usepackage{19lomcon}        % Proceedings volume layout metrics
\usepackage{cite}             % Smart range citations
\usepackage{epsfig}           % Encapsulated PostScript figure inclusion
\usepackage{epstopdf}
\usepackage{subcaption}
\usepackage{amssymb}
\usepackage{amsmath}
\usepackage{bbold}
\usepackage{slashed}
\usepackage{color}

\begin{document}

%%%%%%%%%%%%%%%%%%%%%%%%%%%%%%%%%%%%%%%%%%%%%%%%%%%%%%%%%%%%%%%%%%
% The preamble of the paper
%%%%%%%%%%%%%%%%%%%%%%%%%%%%%%%%%%%%%%%%%%%%%%%%%%%%%%%%%%%%%%%%%%

\title{Reducible contributions to the propagator and EHL in background fields}

\author{N. Ahmadiniaz\,\email{n.ahmadiniaz@hzdr.de (Speaker)}}

\affiliation{Helmholtz-Zentrum Dresden-Rossendorf, Bautzner Landstra\ss e 400, 01328, Dresden, Germany}
\author{F. Bastianelli\,\email{fiorenzo.bastianelli@bo.infn.it}}
\affiliation{Dipartimento di Fisica ed Astronomia, Universit\`a di Bologna and INFN, Sezione di Bologna, Via Irnerio 46, I-40126 Bologna, Italy}
\author{O. Corradini\,\email{olindo.corradini@unimore.it}}
\affiliation{Dipartimento di Scienze Fisiche, Informatiche e Matematiche,
 Universit\`a degli Studi di Modena e Reggio Emilia, Via Campi 213/A, I-41125 Modena and  INFN, Sezione di Bologna, Via Irnerio 46, I-40126 Bologna, Italy}
 \author{J.P. Edwards\,\email{james.p.edwards@umich.mx} and C. Schubert\,\email{schubert@ifm.umich.mx} }
\affiliation{Instituto de F\'isica y Matem\'aticas
Universidad Michoacana de San Nicol\'as de Hidalgo
Edificio C-3, Apdo. Postal 2-82
C.P. 58040, Morelia, Michoac\'an, Mexico}
%\author{A. Ilderton\email{anton.ilderton@plymouth.ac.uk}}
%\affiliation{Center for Mathematical Sciences, University of Plymouth, Plymouth, PL4 8AA, UK}
%\author{C. Schubert\email{schubert@ifm.umich.mx} }
%\affiliation{Instituto de F\'isica y Matem\'aticas
%Universidad Michoacana de San Nicol\'as de Hidalgo
%Edificio C-3, Apdo. Postal 2-82
%C.P. 58040, Morelia, Michoac\'an, Mexico}

% You may repeat \author and \affiliation as many times as necessary!

\date{}
% Print it out!
\maketitle

%%%%%%%%%%%%%%%%%%%%%%%%%%%%%%%%%%%%%%%%%%%%%%%%%%%%%%%%%%%%%%%%%%
% The preamble of the paper
%%%%%%%%%%%%%%%%%%%%%%%%%%%%%%%%%%%%%%%%%%%%%%%%%%%%%%%%%%%%%%%%%%
\bigskip
\begin{abstract}
   We briefly discuss the recent discovery of reducible contributions to QED effective actions due to the presence of external electromagnetic fields at tree level and higher loop-order. We classify the physical effects of these contributions for various field configurations and discuss the strong field asymptotic limit. 
\end{abstract}
\section{Introduction}
\label{intro}
\begin{figure}[h]
    \centering
    \includegraphics[width=0.2\textwidth]{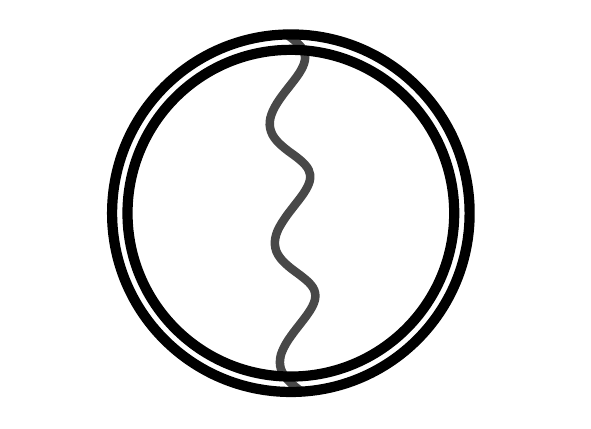}
    \caption{Two-loop irreducible contribution to the EHL.}
    \label{fig1}
\end{figure}

 After the development of quantum mechanics, it has been understood that this theory is not exact and it might be the low-energy limit of a more fundamental quantum theory such as quantum electrodynamics (QED) which was invented in the 30s. After the groundbreaking theory of Dirac \cite{dirac28} and the prediction of the positron, QED became one of the most important and tested theories in science, yet it remains a subject of active investigation. Euler and Kockel \cite{euko35} were the first to study the lowest order corrections to the quantum vacuum. Later Euler and Heisenberg \cite{euhe36} for spinor QED and Weisskopf \cite{weiss36} for scalar QED presented their effective Lagrangians in a classical constant background field for which one can compute the exact one-loop effective action. Their final results are written in a simple closed form corresponding to a sum over an infinite number of Feynman diagrams (with an even number of external photons) which later was confirmed by Schwinger in his proper-time method \cite{schw51}. In the proper-time representation, the renormalized effective Lagrangian to one-loop order obtained by Euler and Heisenberg (EHL) is (in units $\hbar=c=1$)
 \begin{eqnarray}
\hspace{-2.5em}\mathcal{L}^{(1)}_{\rm EH}=-\frac{1}{2}(a^2-b^2)-\frac{1}{8\pi^2}\int_0^\infty \frac{dT}{T^3}e^{-m^2 T}\Big[\frac{(eaT)(ebT)}{\tanh[eaT]\tan[ebT]}-\frac{e^2}{3}(a^2-b^2)T^2-1\Big]
\end{eqnarray}
with two invariants 
 \begin{eqnarray}
 a^2-b^2=-\frac{1}{2} F_{\mu\nu}F^{\mu\nu}\equiv 2\mathcal{F}
  ~~~,~~~ab=-\frac{1}{4}F_{\mu\nu}\tilde{F}^{\mu\nu}\equiv \mathcal{G} 
 \end{eqnarray}
%where the dual tensor is defined as $\tilde{f}^{\mu\nu}=\frac{1}{2} \epsilon^{\mu\nu\alpha\beta}f_{\alpha\beta}$. 
where $a=\sqrt{\sqrt{\mathcal{F}^2+\mathcal{G}^2}-\mathcal{F}}$ and $b=\sqrt{\sqrt{\mathcal{F}^2+\mathcal{G}^2}+\mathcal{F}}$.
The EHL is nonlinear in the background field which indicates new and interesting phenomenological effects like vacuum polarisation, (Sauter-Schwinger) pair production and photon-photon scattering \cite{sauter31,karneu51}, vacuum birefringence \cite{euko35,euhe36}, photon dispersion and photon splitting \cite{adler71,adlsch96} to name a few, see \cite{dunne05} for a comprehensive review. The first radiative corrections to the EHL (see Fig. \ref{fig1}), describing the effect of an additional photon exchange in the loop, were obtained by {\it Ritus} \cite{ritus75}. He obtained $\mathcal{L}^{(2)}_{\rm scal,spin}$ in terms of two-parameter integrals which are intractable analytically; closed-form expressions have been found for their weak-field expansions for the purely electric or magnetic cases in \cite{dunsch00}. \vspace{-0.2cm}
\section{1PR contribution to the EHL and propagators in constant fields }
\begin{figure}[h]
    \centering
    \begin{subfigure}[b]{0.35\textwidth}
        \includegraphics[width=\textwidth]{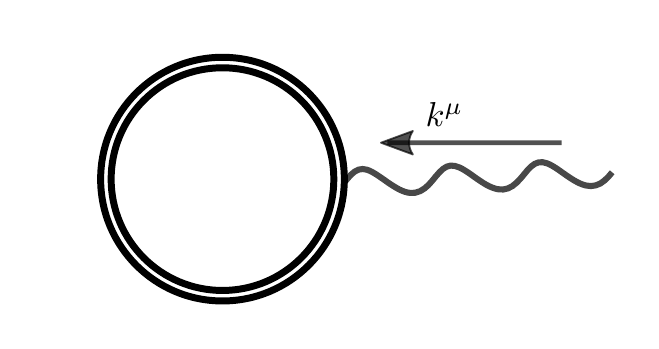}
        \caption{Tadpole diagram}
        \label{fig2}
    \end{subfigure}
    ~~~~~~~~~~~~~~~~~~~~~~~~~~~ %add desired spacing between images, e. g. ~, \quad, \qquad, \hfill etc. 
    \begin{subfigure}[b]{0.35\textwidth}
        \includegraphics[width=\textwidth]{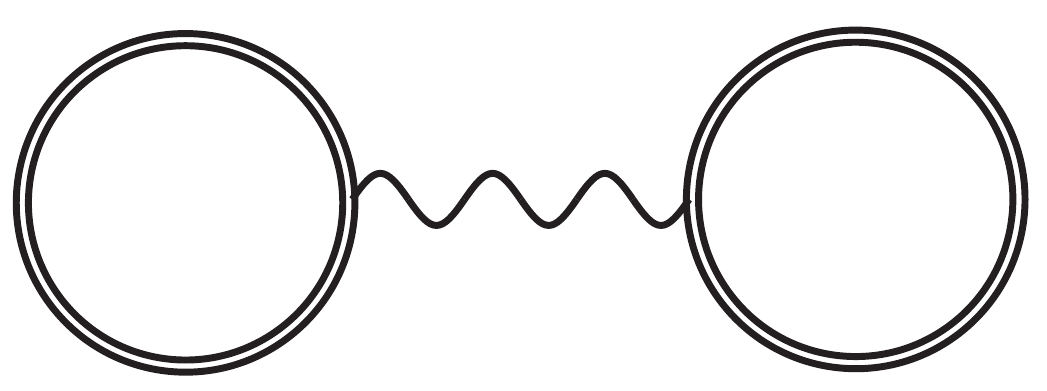}
        \caption{{\rm Dumbbell} diagram}
        \label{fig3}
    \end{subfigure}
    \caption{Tadpole and two-loop 1PR to the EHL.}
    \end{figure}

\begin{figure}[h]
    \centering
    \begin{subfigure}[b]{0.35\textwidth}
        \includegraphics[width=\textwidth]{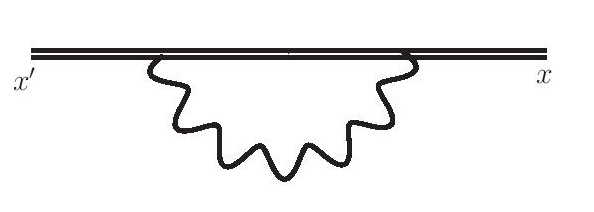}
        \subcaption{1PI}
        \label{fig4}
    \end{subfigure}
    ~~~~~~~~~~~~~~~~~~~~~~~~~~~ %add desired spacing between images, e. g. ~, \quad, \qquad, \hfill etc. 
    \begin{subfigure}[b]{0.35\textwidth}
        \includegraphics[width=\textwidth]{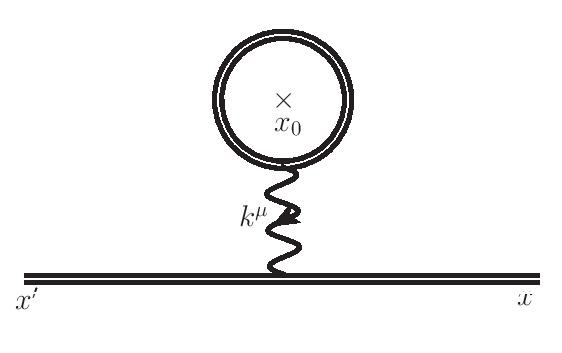}
        \subcaption{1PR}
        \label{fig5}
    \end{subfigure}
    \caption{One-loop correction to the propagator.}
    \end{figure}

 Recently, however, {\it Gies} and {\it Karbstein} \cite{giekarb17} found that historical calculations had overlooked the possibility of \textit{one particle reducible} (1PR) contributions to processes in constant fields. These extra contributions involve a tadpole, displayed in Fig. \ref{fig2}, attached somewhere in a larger Feynman diagram describing the process. The tadpole diagram alone formally vanishes by momentum conservation and gauge invariance, see \cite{ditreu85,ritus75}, but can contribute when sewn to a larger diagram. For example, joining two tadpoles (to make a {\it dumbbell}) in any covariant gauge, leads to a momentum integral of the form \cite{giekarb17} (the $\frac{1}{k^{2}}$ comes from the propagator joining the tadpoles and it is precisely its IR divergence which leads to a non-vanishing result)
 \begin{equation}
 \int d^Dk \,\delta^D(k)\frac{k^\mu k^\nu}{k^2}=\frac{1}{D}\eta^{\mu\nu}
 \label{delta}
 \end{equation}    
 which is the origin of surviving contributions from reducible diagrams. 
\noindent In \cite{giekarb17} it is shown that actually the diagram in Fig. \ref{fig3} does give a finite contribution, given by the following simple formula
\begin{equation}
\mathcal{L}_{\rm EH}^{(2)1{\rm PR}}=\frac{\partial \mathcal{L}_{\rm EH}^{(1)}}{\partial F^{\mu\nu}}\frac{\partial \mathcal{L}_{\rm EH}^{(1)}}{\partial F_{\mu\nu}}=\mathcal{F}\Big[\Big(\frac{\partial\mathcal{L}_{\rm EH}^{(1)}}{\partial\mathcal{F}}\Big)^2-\Big(\frac{\partial\mathcal{L}_{\rm EH}^{(1)}}{\partial\mathcal{G}}\Big)^2\Big]+2\mathcal{G}\frac{\partial\mathcal{L}_{\rm EH}^{(1)}}{\partial\mathcal{F}}\frac{\partial\mathcal{L}_{\rm EH}^{(1)}}{\partial\mathcal{G}}\,.
\end{equation}
Corrections to low energy photon amplitudes arising from these contributions were determined in \cite{AJC}. Later it was then found that there were additional reducible contributions to the scalar and spinor propagators in a constant background field even at the {\it one-loop} order in \cite{edwsch17,ahm17}.  See Fig. (\ref{fig4}) (irreducible) and Fig. (\ref{fig5}) (reducible) for one-loop corrections to the scalar/fermion propagator in a constant field. 
\vspace{-0.2cm} 
 In \cite{edwsch17}, the authors used a direct calculation of all the ingredients,
using the worldline approach to QED in a constant field  \cite{strassler,shaisultanov,adlsch96,40,review, reviewcj, ahm17}. Here, as in \cite{ahm17}, we will proceed more efficiently: rather than actually calculating the one-photon spinor propagator in the field, we will write down its worldline path integral representation, and manipulate it to show that its linear part in the photon momentum $k^{\mu}$ -- which is all that is required for the sewing -- satisfies (with the background field in Fock-Schwinger gauge)
\begin{eqnarray}
S^{x'x}_{(1)}\Bigl\vert_k 
%&=&-2i\varepsilon \cdot \left(\frac{\partial S^{x'x}}{\partial F} + \frac{ie}{2}x_- S^{x'x} x_+ \right) \cdot k -\frac{ie}{2} \varepsilon\cdot \big(\gamma x_- +x_-\gamma \big)\cdot k K^{x'x}
 %\nonumber\\
&=& -2i\varepsilon \cdot \left(\frac{\partial S^{x'x}}{\partial F} + \frac{ie}{2}x^{\prime} S^{x'x} x\right) \cdot k  +\varepsilon\cdot L \cdot k
\label{idopen}	
\end{eqnarray}
where $x_+ \equiv \frac{1}{2} (x+x')$, $x_- \equiv x' -x$, and
$L^{\mu\nu}$ is a symmetric tensor that will not contribute to the sewing due to symmetry. Together with the similar identity for the closed loop \cite{edwsch17},
\begin{equation}
\Gamma^{(1)}_{(1)}[k,\varepsilon;F] 
= -2i (2\pi)^{D} \delta^D(k) 
\Bigl\lbrack\varepsilon\cdot\frac{\partial   {\cal L}^{(1)} (F)}{\partial F} \cdot k + O(k^3)\Bigr\rbrack\, 
\label{idclosed}
\end{equation}
it is easy to go for the definition for the 1PR part of the self-energy
\begin{equation}
S^{x'x(1)1PR}
= \int \frac{d^Dk}{(2\pi)^D k^2} 
\Gamma^{(1)}_{(1)}[k',\varepsilon';F]S^{x'x}_{(1)}[k,\varepsilon;F]
\Big\vert_{k' \rightarrow-k,\,\varepsilon^{\mu}\varepsilon'^{\nu} \rightarrow\eta^{\mu\nu}}  
\label{sewingGammaD}
\end{equation}
and (\ref{delta}) to obtain for spinor QED 
\begin{equation}
S^{x'x(1)1PR} = 	\frac{\partial S^{x'x}}{\partial F_{\mu\nu}}\frac{\partial \mathcal{L}^{(1)}}{\partial F^{\mu\nu}} 
+  \frac{ie}{2}S^{x'x}x^{\prime\mu}\frac{\partial \mathcal{L}^{(1)}}{\partial F^{\mu\nu}}x^{ \nu} 
	\label{resCompspin}
\end{equation}
and, by Fourier transformation, the momentum space version thereof,
\begin{equation}
S^{(1)1PR}(p)  = \frac{\partial S(p\,\vert F)}{\partial F_{\mu\nu}}\frac{\partial \mathcal{L}^{(1)}}{\partial F^{\mu\nu}}  
	\label{resComppspin}
\end{equation}
 where $S(p\vert F)$ is the free propagator in the constant field, see \cite{ahmad17} for the scalar and \cite{114} for the spinor case. The proof of this so called  ``derivative identity'' can be found in detail in \cite{ahm17} but to obtain the tadpole in Fig. \ref{fig4} we need these identities, along with the one-loop effective action and the propagator in a general constant field background to be discussed in the following.   \par 
 %\subsection{The closed loop with zero and one photons}
   The worldline representation of the one-loop spinor QED effective action can be written as
a double worldline path integral (see \cite{review, reviewcj} and refs. therein):
\begin{eqnarray}
&&\Gamma^{(1)}
 [A]  = - \frac{1}{2}\int_0^\infty 
\frac{dT}{T}
%{\lbrack 4\pi T\rbrack}^{-{D\over 2}}
e^{-m^2T}
\int_P D x
\int_A 
D\psi\nonumber\\
& \phantom{=}
&\times
{\rm exp}\biggl [- \int_0^T d\tau
\Bigl (\frac{1}{4}{\dot x}^2 + {1\over
2}\psi\cdot\dot\psi
+ ieA\cdot\dot x - ie\, \psi\cdot F\cdot\psi
\Bigr )\biggr ]
\, .
\label{spinorpi}
\end{eqnarray}
Here the orbital path integral $\int D x$ is over closed trajectories in space-time, $x(T) = x(0)$,
the spin path integral $\int D\psi$ over Grassmann functions $\psi^{\mu}(\tau)$
obeying antiperiodic boundary conditions, $\psi^{\mu}(T) = - \psi^{\mu}(0)$. 
For a constant $F_{\mu\nu}$ it is convenient to use Fock-Schwinger gauge centered at the loop center of mass $x_0$, 
since this allows one to write $A_{\mu}$ in terms of $F_{\mu\nu}$:
\begin{eqnarray}
A_{\mu}(x) = \frac{1}{2} (x-x_0)^{\nu}F_{\nu\mu} \, .
\label{FS}
\end{eqnarray}
Separating off the loop center of mass via $x(\tau) = x_0 + q(\tau)$,
one obtains 
\begin{eqnarray}
\Gamma^{(1)}(F) = \int d^D x_0 {\cal L}^{(1)} (F) 
\label{GammatoL}
\end{eqnarray}
where
\begin{eqnarray}
&&{\cal L}^{(1)} (F) 
= - \frac{1}{2}\int_0^{\infty}
\frac{dT}{T}
%{\lbrack 4\pi T\rbrack}^{-{D\over 2}}
e^{-m^2T}
\int _P 
D q
\int _A 
D\psi\nonumber\\
& \phantom{=}
&\times
{\rm exp}\biggl [- \int_0^T d\tau
\Bigl (\frac{1}{4}{\dot q}^2 +\frac{1}{2}\psi\cdot\dot\psi
+ \frac{ie}{2} q\cdot  F\cdot\dot q - ie \,\psi\cdot F\cdot\psi
\Bigr )\biggr ]
\label{spinorpiq}
\end{eqnarray}
and $q^{\mu}(\tau)$ now obeys the ``string-inspired'' constraint $\int_0^Td\tau q^{\mu}(\tau) =0$. This can be written as \cite{review,18}
 \begin{equation}
{\cal L}^{(1)}(F) = 
-2\int_0^{\infty}\frac {dT}{T}
(4\pi T)^{-\frac{D}{2}}
e^{-m^2T}
{\rm det}^{-\frac{1}{2}}
\biggl[{{\rm tan}{\cal Z}\over {\cal Z}}\biggr]
\label{L1spinF}
\end{equation}
which is a suitable representation of the one-loop EHL for our purpose here. \newline Next, we need an analog representation for the free propagator in a constant field. 
%To build the reducible diagram in Fig. \ref{fig5} in one side we need the one-loop one-photon amplitude $\Gamma^{(1)}_{(1)}[k,\varepsilon;F]$ which is obtained from (\ref{spinorpi}) by the insertion of the photon vertex operator
%\begin{eqnarray}
%-ieV[k,\varepsilon]
%= -ie e^{ik\cdot x_0} 
%\int_0^Td\tau
%\Bigl[
%\varepsilon\cdot \dot q
%+2i
%\varepsilon\cdot\psi
%k\cdot\psi
%\Bigr]
%\,e^{ik\cdot q}
%\nonumber\\
%\label{photonvertop}
%\end{eqnarray}
%under the path integrals in (\ref{spinorpiq}). Since we are interested only in the part of this amplitude linear in the photon momentum $k$ we observe that, when acting on the exponent in (\ref{spinorpiq}), we can further replace the integrand in this last expression under the path integral by
%\begin{eqnarray}
%V[k,\varepsilon]\Big\vert_k \to 
%-2i e^{ik\cdot x_0}  \varepsilon^{\mu}\frac{\partial}{\partial F^{\mu\nu}} k^{\nu} .
%\end{eqnarray}
%Putting things together we obtain
%\begin{eqnarray}
%\Gamma^{(1)}_{(1)}[k,\varepsilon;F] 
%= -2i (2\pi)^{D} \delta^D(k) \varepsilon\cdot\frac{\partial}{\partial F} \cdot k \, {\cal L}^{(1)} (F).
%\label{olop}
%\end{eqnarray}
%\subsection{The open line with zero and one photons}
An efficient worldline representation of the dressed spinor propagator in a constant field has been developed only very recently \cite{114}, based on
\cite{fragit,18}. Without photons it leads to the representation
\footnote{Our conventions follow \cite{srednicki-book}, except for the opposite sign of the elementary charge.} 
\begin{eqnarray}
	S^{x'x}(F)  &=& \left[m + i \gamma \cdot \left(\frac{\partial}{\partial x'} - \frac{ie}{2}F \cdot x_-\right)\right]
	K^{x'x}(F) 
	\label{D0photon}
\end{eqnarray}
where $x_- \equiv x' -x$ with the ``kernel'' function
\begin{eqnarray}
	\hspace{-1.5em} K^{x'x}(F)
%& =&\int_0^\infty e^{-m^2 T} dT\int_{x(0)=x}^{x(T)=x'}Dx\, e^{-\int_0^Td\tau \big (\frac{1}{4}{\dot x}^2 
%+ ieA\cdot\dot x \big)   }\nonumber\\
%&& \quad  \times \frac{1}{4} {\rm symb}^{-1}\Biggl\{\int_{A}D\psi\, e^{-\int_0^Td\tau \big [ {1\over
%2}\psi\cdot\dot\psi
 %- ie\, \big(\psi+\frac{1}{2} \eta\big)\cdot F\cdot\big(\psi +\frac{1}{2} \eta\big)
%\big]} \Biggr\} \nonumber\\
&=& \int_{0}^{\infty} dT e^{-m^{2}T} (4\pi T)^{-\frac{D}{2}} {\rm det}^{-\frac{1}{2}}\Big[\frac{\tan\cal{Z}}{\cal Z}\Big]
	e^{-\frac{1}{4T} x_- \cdot {\cal{Z}} \cdot \cot {\cal{Z}} \cdot {x_-}}\, \nonumber\\
	&& \hspace{3cm}\times{\rm symb}^{-1}\Bigl[e^{\frac{i}{4} \eta \cdot \tan \cal{Z} \cdot \eta}\Bigr]\,.
\label{K0photon}
\end{eqnarray}
Here the propagation is from $x$ to $x'$, ${\cal Z}_{\mu\nu} \equiv eF_{\mu\nu} T$ and we used Fock-Schwinger gauge at $x$. 
The constant Grassmann vector $\eta$ generates the $\gamma$-matrix structure of the propagator via the ``symbol map''  defined by 
$
{\rm symb} 
\bigl(\hat\gamma^{[\alpha\beta\cdots\rho]}\bigr) \equiv 
\eta^\alpha\eta^\beta\ldots\eta^\rho
$
where $\hat\gamma^{\mu} \equiv i\sqrt{2} \gamma^{\mu}$
and $\hat\gamma^{[\alpha\beta\cdots\rho]}$ denotes the totally antisymmetrized product.
Now, with (\ref{L1spinF}) and (\ref{K0photon}) we can build the tadpole diagram in Fig. \ref{fig4} using the above ``derivative identity.'' In configuration space, an explicit representation of the addendum can be found by simply carrying out the differentiation (all matrices are built from the constant anti-symmetric field strength tensor and so commute) of the un-dressed propagator and one-loop effective action \cite{ahm17}. Here we just quote the final expression in momentum space which is obtained by Fourier transforming the $x$-space result
\begin{eqnarray}
\hspace{-2.5em}S^{(1)1PR}(p)&=&e^{2}\int_{0}^{\infty} dT T\,e^{-T(m^{2} + p \cdot \frac{\tan {\cal Z}}{{\cal Z}} \cdot p)}\int_{0}^{\infty}dT^{\prime}(4\pi T^{\prime})^{-\frac{D}{2}}e^{-m^{2}T^{\prime}}
\textrm{det}^{-\frac{1}{2}}\Bigl\lbrack\frac{\tan {\cal Z}'}{{\cal Z}'} \Bigr\rbrack\nonumber \\
\hspace{-2.5em}&&\times \bigg\lbrace\Bigl\lbrack m - \gamma \cdot(\mathbb{1} + i{\tan{\cal Z}}) \cdot p  \Bigr\rbrack
\bigg[- T p \cdot \frac{{\cal Z} - \sin {\cal Z} \cdot \cos {\cal Z}}{{\cal Z}^{2} \cdot \cos^{2}{\cal Z}} \cdot \Xi^{\prime} \cdot p +  \Xi^{\prime}_{\mu\nu}\frac{\partial}{\partial {\cal Z}_{\mu\nu}} \bigg] 
\nonumber\\
\hspace{-2.5em}&&\,\,\, - i \gamma \cdot \sec^{2}{\cal Z} \cdot
\Xi^{\prime} \cdot p 
\bigg\rbrace
\textrm{symb}^{-1} \bigg\{ e^{\frac{i}{4}\eta \cdot \tan {\cal Z} \cdot \eta}\bigg\} 
\label{S1}
\end{eqnarray}
where $
\Xi=\frac{d}{d{\cal Z}}\ln\Big[\frac{\tan {\cal Z}}{\cal Z}\Big]$ see \cite{ahm17}.
\section{Pure magnetic field background}
    In this section we consider a pure constant magnetic field background pointing along the $z$-direction, say, so that $\vec{B}=B\hat{z}$. In this background there are only two non-vanishing components of the field strength tensor,  $F_{12}=-B$ and $F_{21}=B$. In the following two subsections we compute the 1PR contribution to the two-loop EHL and the tadpole diagram in a magnetic field background. 
    \subsection{1PR to the two-loop EHL}
    The two-loop EHL (spinor case) in this background has the following parameter integral representation 
    \begin{eqnarray}
    \mathcal{L}^{(2)1PR}=\frac{4e^2}{D}\int_0^\infty ds(4\pi is)^{-\frac{D}{2}}\mathcal{J}(z)\int_0^\infty ds'(4\pi i s')^{-\frac{D}{2}}\mathcal{J}(z')
    \label{eq1}
    \end{eqnarray}
    where $\mathcal{J}(z)=(z/\tan z)(\cot z-1/z+\tan z)$ with $z=eBs$. Note that the integrand contains arbitrary positive powers of $z$ and $z'$ therefore it is clear that this contribution to the EHL cannot be absorbed by renormalization, so it implies important physical corrections at two-loop \cite{ahm19}.  For weak magnetic fields one can expand the integrand and compute the first non-trivial contributions. After the expansion and performing the parameter integrals in (\ref{eq1}) 
  \begin{eqnarray}
     \mathcal{L}^{(2)1PR}=\frac{4e^2}{D m^4}\,(\frac{m^2}{4\pi})^D\,\Big[\frac{2}{3}\,(\frac{eB}{m^2})\,\Gamma[2-\frac{D}{2}]-\frac{4}{45}\,(\frac{eB}{m^2})^3\,\Gamma[4-\frac{D}{2}]+\cdots\Big]^2.
  \end{eqnarray}
In $D=4$ the first term of the above expansion is divergent but still linear in the coupling of the respective loop to the background field ($eB$) so it can be removed by renormalization, but higher order terms are finite and physical so they should be contrasted with the weak field expansion of the irreducible contribution to the two-loop EHL \cite{ritus752}. The opposite strong-field limit is also interesting. Since there are two copies of the same integral in (\ref{eq1}) it suffices to focus on one and extract the strong limit behaviour. It has recently been shown by {\it Karbstein} in \cite{karb19} that the strong field asymptotic limit of EHL is determined (at all loop orders) by the {\it reducible} contributions. After renormalisation the parameter integrals are evaluated and setting $D=4-2\epsilon$, the pole cancels out between difference pieces of the integral in (\ref{eq1}): the remaining, finite, expression can then be expanded for large $B$ to obtain the leading order behaviour
\begin{eqnarray}
\mathcal{L}^{(2)1PR}\sim \frac{1}{2} B^2\Big[\alpha\beta_1\ln \Big(\frac{eB}{m^2}\Big)\Big]^2\,; \qquad \frac{eB}{m^{2}} \gg 1
\end{eqnarray}
where $\beta_1=1/3\pi$ is the order $\alpha$ coefficient of the $\beta$-function in spinor QED -- see \cite{ahm19} for details. It confirms the results given in \cite{karb19} at two-loop order. 
%\begin{figure}[h]
   % \centering
    %\includegraphics[width=0.3\textwidth]{fig-tadpole.pdf}
    %\caption{Tadpole diagram in a constant background field.}
    %\label{fig2}
%\end{figure}
\subsection{1PR to the tadpole}
As for the 1PR contribution to the spinor self-energy, after plugging the field strength tensor into (\ref{S1}) we find that the one-loop 1PR correction in a constant magnetic field is given by 
\begin{eqnarray}\label{igen}
\hspace{-4em}	&&S^{(1)1PR}(p) = -ie^{2}\int_{0}^{\infty} \, ds\, s\,e^{-is(m^{2}+p_\parallel^2 + \frac{\tan z}{z}p_\perp^{2})}\int_{0}^{\infty}\, ds^{\prime}(4i \pi s^{\prime})^{-\frac{D}{2}}e^{-im^{2}s^{\prime}}
\big(\cot z'-z'\csc^2 z'\big) \nonumber \\
\hspace{-4em}&&\hspace{0.7cm}\times\bigg\lbrace\Big[-is p_{\perp}^{2}\Big(\frac{\sec^2z}{z}-\frac{\tan z}{z^2}\Big)\big( m-\slashed{p}+{\rm tan}z\, \gamma^{[2}p^{1]}\big) +\sec^2 z\, \gamma^{[2}p^{1]} \Big] \Big[\mathbb{1}+\frac{1}{2}\tan z [\gamma^{2}, \gamma^{1}]\Big] \nonumber \\
\hspace{-4em} &&\hspace{4cm}+\frac{1}{2}(m-\slashed{p}+\tan z\,\gamma^{[2}p^{1]})\sec^{2}z[\gamma^{2}, \gamma^{1}]\bigg\}
\end{eqnarray}
where $z=eBs$, $p_\parallel=(p_0,0,0,p_3)$ and $p_\perp=(0,p_1,p_2,0)$. This general result is non-vanishing, as we show directly below and generalises easily to a constant magnetic field in an arbitrary direction. The integrand involves arbitrary powers of $z'$, so cannot be completely absorbed by renormalisation. The parameter integrals in (\ref{igen}) may be done numerically. To cubic order in the magnetic field the $s'$ integral provides a factor
\vspace{-0.2cm}\begin{figure}[h]
	\includegraphics[width=1\textwidth]{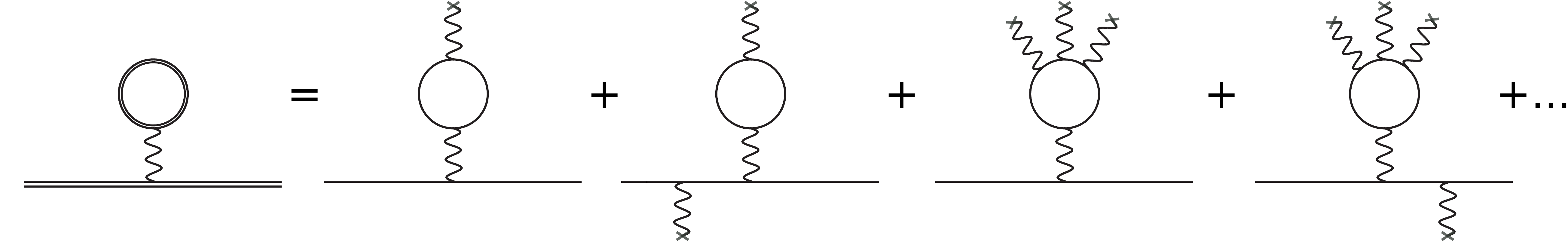}
	\caption{\label{fig-expansion} Diagrammatic representation of the weak-field expansion for the 1PR contribution to the electron propagator in a constant magnetic field. }
	%An odd number of external photons are attached to the loop (due to Furry's theorem) and an arbitrary number can be attached to the line.}
	\end{figure}
\begin{eqnarray}
	&&\int_{0}^{\infty}ds^{\prime}  \Big(4i \pi s^{\prime}\Big)^{-\frac{D}{2}} e^{-im^{2}s^{\prime}}
\Big(\cot z'-z'\csc^2 z'\Big)=\frac{2}{3 m^{2}} \left(\frac{ m^{2}}{4\pi}\right)^{\frac{D}{2}}\nonumber
\\
&&\hspace{1cm}\times\bigg\{\left(\frac{eB}{m^{2}}\right) \Gamma\Big[2 - \frac{D}{2}\Big] - \left(\frac{eB}{m^{2}}\right)^{3} \frac{2}{15} \Gamma\Big[4 - \frac{D}{2}\Big]  + \ldots \bigg\}\,.
\end{eqnarray}
The next step is to expand the $s$-integrand in $z$. The remaining proper time integral over $s$ then yields
\begin{eqnarray}
	&&S^{(1)1PR}(p)\approx\frac{2ie^{2}}{3} \left(\frac{m^{2}}{4 \pi}\right)^{\frac{D}{2}}\left[ \frac{eB}{m^2}\Gamma\Big[2 - \frac{D}{2}\Big] - \frac{2}{15} (\frac{eB}{m^2})^{3}  \Gamma\Big[4 - \frac{D}{2}\Big]+ \ldots \right]\nonumber\\
	&&\times\bigg[ \frac{1}{4 m^{2}} \frac{ \big\{ m - \slashed{p}\,, [\gamma^{2}, \gamma^{1}] \big\}}{(m^{2} + p^{2})^{2} } +4i \left(\frac{eB}{m^{2}}\right) \left(  \frac{(m-\slashed{p})p_{\perp}^{2}}{(m^{2} + p^{2})^{4}} -  \frac{p^{1}\gamma^{1} + p^{2}\gamma^{2}}{(m^{2} + p^{2})^{3}}\right)  + \ldots \bigg] \nonumber\\
	\label{S1B}
\end{eqnarray}
which is also represented diagrammatically in Fig. \ref{fig-expansion}.  Here the top line is the contribution from the loop; the first term in square brackets diverges in $D = 4$. However, being linear in the coupling to the background, this can be absorbed by a renormalisation. The first non-trivial contribution is of order $(eB/m^2)^3$ which is clearly finite. In the next section we briefly discuss the case of constant crossed field and plane wave background fields. 
\section{Constant crossed field and plane wave backgrounds}
We consider the class of constant fields with vanishing Maxwell invariants, $F_{\mu\nu}F^{\mu\nu} =0$ and $F_{\mu\nu}\widetilde{F}^{\mu\nu} =0$, where the dual field strength tensor is defined as usual by $\widetilde{F}^{\mu\nu} := \frac{1}{2}\epsilon^{\mu\nu\alpha\beta}F_{\alpha\beta}$. Furthermore $F^{3}_{\mu\nu} \equiv F_{\mu\alpha} F^{\alpha\beta} F_{\beta\nu} = 0$ for such fields and all higher powers also vanish. For constant crossed fields we may always choose coordinates such that the field strength tensor has $F_{01}=F_{13}=B$. % It can be checked that for such a field strength tensor $F_{\mu\nu}^3=0$, so that all hyperbolic trigonometric functions that enter the proper time representations of the general 1PR contributions, above, are at most quadratic in ${\cal Z}$ and ${\cal Z}'$. 
 For the 1PR contribution to the self-energy, evaluating the trigonometric functions in (\ref{S1}) and computing the $s'$ integral leads to the representation (the super-script minus refers to light-cone coordinates, $x^\pm:=x^0\pm x^3$ and the square bracket anti-symmetrization of indices without combinatorical factor)
\begin{eqnarray}
\label{Scd}\hspace{-2.5em}&&S^{(1)1PR} = \frac{e^{2}}{m^{2}}\left(\frac{m^{2}}{4 \pi}\right)^{\frac{D}{2}} \frac{eB}{m^{2}}\Gamma\Big[2 - \frac{D}{2}\Big]\int_{0}^{\infty} d s \,s\,e^{-is\left(p^{2} + m^{2} + \frac{z^{2}}{3} {p^-}^{2}\right)} \nonumber\\
\hspace{-2.5em}&& \left[ \frac{1}{2}\Big\{ i\left(m - \slashed{p}\right) \, , \, \frac{4is}{9} z {p^-}^{2} - \frac{2}{3} \gamma^{-}\gamma^{1} \Big\} + iz\gamma^{[-}p^{1]} \left(\frac{4is}{9} z {p^-}^{2} - \frac{2}{3} \gamma^{-}\gamma^{1}\right)\right] \Big[\mathbb{1} + z\gamma^{-}\gamma^{1} \Big]\nonumber\\
\end{eqnarray}
in which $\{\cdot, \cdot \}$ denotes the anticommutator, and $z = eBs$. 
%The leading $\frac{eB}{m^2}$ arises from the integral over the loop proper time $s^{\prime}$ and, as argued above, the result is linear in this coupling of the loop to the crossed field background (this is because $\Xi'$ is linear in ${\cal Z}'$ and it enters every term of the integrand). As such we see that this 1PR contribution can be absorbed simply by an additional (infinite, in $D = 4$) renormalisation of the photon propagator, see \cite{ahm19} for more detail. It therefore has no physical significance, once the photon propagator has been correctly renormalised.
The same analysis can be done for the plane wave background which we will not present here, see \cite{ahm19}. 
%Here the logic is somewhat reversed: there is no a priori reason to expect that tadpole contributions vanish, yet one would like to show that they do not change the vanishing of the effective action in a plane wave background. 
As the constant crossed field case the plane wave background does not lead to any finite contribution for the self-energy diagram due to insufficient Lorentz invariant quantities that can be constructed from the physical parameters of the problem, see \cite{ahm19}. In \cite{ahm19} we considered background plane waves of arbitrary strength and shape. Here we were able to make a stronger statement; we calculated the 1PR correction to \textit{any} diagram, and showed that this again amounts to a divergence (in $D = 4$) which can be renormalised away. This is consistent with, and goes beyond, one-loop Hamiltonian-picture calculations where the tadpole does not appear due to normal ordering~\cite{Ilderton:2013dba}, as in background-free QED.  For all plane waves, including constant crossed fields, we have also confirmed that the dumbbell diagram vanishes identically. Therefore (unlike in the case of magnetic fields) we confirm there is no additional two-loop correction to the effective action coming from the 1PR diagrams. 
\section{Conclusion}
We supplied a brief review on the recently discovered 1PR contribution to QED in the presence of constant background fields. We consider a general electromagnetic field, a pure magnetic field, constant crossed and plane wave background fields. One obtains a finite contribution for a general background, which in the case of a pure magnetic field leads to very interesting results in both strong and weak field limits. For the crossed and plane wave cases we show these 1PR contributions are linear in the background field and can therefore be removed by renormalization which indicates that 1PR does not lead to any physical contribution. For the plane wave case we make the stronger statement that 1PR corrections to any diagram can be renormalized away. \bigskip 

\textbf{Acknowledgements:} %NA would like to thank the organizers of the Nineteenth Lomonosov Conference on the Elementary Particle Physics, Moscow, August 22-28, 2019. 
JPE is pleased to thank the Royal Society for funding through Newton Mobility Grant NMG\textbackslash R1\textbackslash 180368.

%%%%%%%%%%%%%%%%%%%%%%%%%%%%%%%%%%%%%%%%%%%%%%%%%%%%%%%%%%%%%%%%%%
% References
%%%%%%%%%%%%%%%%%%%%%%%%%%%%%%%%%%%%%%%%%%%%%%%%%%%%%%%%%%%%%%%%%%


\begin{thebibliography}{99}
\bibitem{dirac28}
P. A. M. Dirac, Proc. R. Soc. London Ser. A {\bf 117}, 610 (1928).
\bibitem{euko35}
H. Euler and B. Kockel, Naturwissensechaften {\bf 23} 246 (1935).
\bibitem{euhe36}
H. Euler and W. Heisenberg, Z. Phys. {\bf 98}, 714 (1936).  
\bibitem{weiss36}
V. Weisskopf, Kong. \! Dans. Vid. Selsk, Math-fys. Medd. XIV {\bf 6} (1936).
\bibitem{schw51}
J. Schwinger, Phys.Rev. {\bf 82}, 664 (1951). 
\bibitem{karneu51}
R. Karplus and M. Neuman, Phys. Rev. {\bf 83}, 776 (1951).
\bibitem{sauter31}
F. Sauter, Z. Phys. {\bf 82}, 742 (1931).
\bibitem{adler71}
S. Adler, Ann of Phys. {\bf 67}, 599 (1971).
\bibitem{adlsch96}
S. Adler and C. Schubert, Phys. Rev. Lett. {\bf 77} (1996) 1695.
\bibitem{dunne05}
G. V. Dunne, arXiv: 0406216 [hep-th]. 
\bibitem{ritus75}
V. I. Ritus, Sov. Phys. JETP {\bf42} (1975) 774.
\bibitem{dunsch00}
G. V. Dunne, A. Huet and C. Schubert, JHEP {\bf 013} (2006) 0611.
\bibitem{giekarb17}
H. Gies and F. Karbstein, JHEP {\bf 03} (2017) 108.

\bibitem{ditreu85}
W. Dittrich and M. Reuter, Lect. Notes Phys. 220, 1 (1985).
\bibitem{AJC}
  J.~P.~Edwards, A.~Huet and C.~Schubert,
  %``On the low-energy limit of the QED N-photon amplitudes: part 2,''
  Nucl.\ Phys.\ B {\bf 935} (2018) 198
\bibitem{edwsch17}
J. P. Edwards and C. Schubert, Nucl. Phys. {\bf B 923} (2017) 339.
\bibitem{ahm17}
N. Ahmadiniaz, F. Bastianelli, O. Corradini, J. P. Edwards and C. Schubert, Nucl. Phys. {\bf B 924} (2017) 377.
\bibitem{review}
C. Schubert, 
%``Perturbative quantum field theory in the string-inspired formalism'', 
Phys. Rept. {\bf 355}, 73 (2001).  
\bibitem{reviewcj}
J. P. Edwards and C. Schubert, 
%``Perturbative quantum field theory in the string-inspired formalism'', 
arXiv:1912.10004 [hep-th]

\bibitem{strassler}
M. J. Strassler, 
%``Field theory without Feynman diagrams: One-loop effective actions'', 
Nucl. Phys. {\bf B 385} (1992) 145.
\bibitem{shaisultanov}
R. Zh. Shaisultanov, 
%``On the string-inspired approach to QED in external field'', 
Phys. Lett. {\bf B 378} (1996) 354.
%\bibitem{17}
%S.L. Adler, C. Schubert, 
%``Photon splitting in a strong magnetic field: Recalculation and
%comparison with previous calculations'',
%Phys. Rev. Lett. {\bf 77} (1996) 1695.

\bibitem{40} 
C. Schubert,
 %``Vacuum polarisation tensors in constant electromagnetic fields: Part I'', 
Nucl. Phys. {\bf B 585}, 407 (2000).
\bibitem{ahmad17}
A. Ahmad, N. Ahmadiniaz, O. Corradini, S. P. Kim and C. Schubert, Nucl. Phys. {\bf B} 919 (2017) 9.
\bibitem{114}
N. Ahmadiniaz, V. Banda, F. Bastianelli, O. Corradini, J.P. Edwards and C. Schubert, in preparation. 
\bibitem{18}
M. Reuter, M. G. Schmidt and C. Schubert, 
%``Constant external fields ingauge theory and the spin 0, 1/2, 1 path integrals'',  
Ann. Phys. {\bf 259} (1997) 313.

\bibitem{fragit}
E.S. Fradkin, D.M. Gitman,
%``Path integral representation for the relativistic particle propagators and BFV quantization'',
Phys. Rev. {\bf D 44} (1991) 3230.
\bibitem{srednicki-book}
M. Srednicki, {\it Quantum Field Theory}, Cambridge University Press 2007.
 \bibitem{ahm19}
 N. Ahmadiniaz, J. P. Edwards and A. Ilderton, JHEP {\bf 05} (2019) 038.
  \bibitem{ritus752}
V. I. Ritus, Zh. Eksp. Teor. Fiz  {\bf69} (1975) 1517.
\bibitem{karb19}
F. Karbstein, Phys. Rev. Lett. {\bf 122} (2019) 211602.



\bibitem{Ilderton:2013dba}
A.~Ilderton and G.~Torgrimsson, %\emph{{Radiation reaction from QED: lightfront
  %perturbation theory in a plane wave background}},
  %\href{https://doi.org/10.1103/PhysRevD.88.025021}
  Phys. Rev. D {\bf88} (2013) 025021.
  
\end{thebibliography}
\end{document}